**Self-propulsion of a light-powered microscopic crystalline flapper in water**

Kazuma Obara, Yoshiyuki Kageyama,* and Sadamu Takeda

**Abstract**

A key goal in developing molecular microrobots that mimic real-world animal dynamic behavior is to understand better the self-continuous progressive motion resulting from collective molecular transformation. This study reports, for the first time, the experimental realization of directional swimming of a microcrystal that exhibits self-continuous reciprocating motion in a two-dimensional water tank. Although the reciprocal flip motion of the crystals was like that of a fish wagging its tail fin, many of the crystals swam in the opposite direction to which a fish would swim. Here we explore the directionality generation mechanism and physical features of the swimming behavior by constructing a mathematical model for the crystalline flapper. The results show that a tiny crystal with a less-deformable part in its flip fin exhibits a pull-type stroke swimming, while a crystal with a fin that uniformly deforms exhibits push-type kicking motion.

**Introduction**

The development of molecular microrobots with motional dynamics based on the characteristics of living systems has attracted considerable attention from physicists and chemists, and offers enormous potential in applications ranging from intelligent machines to biomedical devices.[1-4] The first challenge to overcome to enable progress toward these potential applications is attaining self-organized macroscopic dynamics in molecular machines. The next challenge is creating directional motion that can be regarded as mechanical work created from the self-sustained transformation of a molecular machine. Generally, the motion of an inanimate object is passive: an externally supplied force or an alternation of environment drives its repetitive movement.[5-9] Only objects possessing an ability to convert supplied energy to kinetic energy on a scale larger than thermal fluctuation can move self-sustainably.[3, 10-11] The directions of its motion are defined by asymmetries of the object or anisotropies in its surroundings, or both. For example, Janus particles, whose surfaces have two distinct chemical or photocatalytic properties, show self-propulsion in aqueous solutions.[12-14] An ∩-shaped chemical-oscillatory hydrogel can display walking motions on a structurally-asymmetric substrate,[15] and a polymer material placed in a rectangular plastic frame can crawl guided by a light source.[16] As can be understood from these examples, self-continuous features and spatial asymmetry are required for matter to show dynamic activity.[17]

Previously, we reported our realization of light-powered self-oscillatory flipping of submillimeter-sized crystals wherein their spatially-and-temporally self-patterned features were characterized not by the light-source direction but by internal chemical and physical processes.[18] The crystal was prepared from a synthesized azobenzene derivative (6-[4-(4-butylphenylazo)phenoxy]hexanoic acid, **1**)[19] and oleic acid. The self-repetitive bending-and-flattening motion occurred via crystalline phase transition triggered by photoisomerization. This powerful autonomous flipping allowed the crystal to exert mechanical force on its surroundings.. In this study, we report the swimming mechanics of tiny self-flipping crystals in a flat-box water pool.

In liquids, such as water, viscous drag forces act against the motion of small objects.[4, 20] This drag causes both a gain in the swimming thrust force as well as prohibition of the inertial motion of the object. As a result, the object stops immediately upon changing its position. If the deformation of an object is a spatially reciprocal process, the sum of the displacement of the object in a Newtonian fluid (in which fluid inertia is negligible) is zero. This feature is known as the scallop theorem.[21]

According to theoretical modeling studies, the scallop theorem can be violated when reciprocal objects deform asymmetrically or when the objects exist under heterogeneous conditions.[22-24] In the present experiments, the crystal displayed reciprocal motion yet showed a net directional movement. This phenomenon suggests that the reciprocal motion was asymmetric on a macroscopic scale. Because the crystals' vertical movement was limited by the flat-box shape of the water pool, we constructed a model in which vertical motion was restricted in order to elucidate the characteristics of crystal swimming motion. Through mathematical analysis employing a simple model of the flipping dynamics, it is revealed that a slight difference in shape determines the swimming direction of the crystal.

**Experimental Section**

General considerations

All experimental observations were obtained as described in our previous paper.[18] A mixture of **1** and sodium oleate was dispersed in phosphate buffered saline (pH 7.5) and ultrasonicated at 55 °C for 30 min, then was cooled to room temperature in darkness. The dispersed crystals were placed into the water pool, which was built using two glass slides (Matsunami Glass, Japan) and a 0.3 mm thick plastic frame (Frame Seal, Bio-Rad, USA). The dynamics of the crystal were recorded using a differential interference contrast microscope (Nikon, Ti-U or TE2000, Japan) equipped with an incident fluorescent unit, the light source of which was a high-pressure mercury lamp, and a CCD camera (Omron Sentech, TC152USB-AH or MCE132U3V, Japan). For measuring outlines and tracking movements of the crystals, we used Image-Pro Premier software, versions 9.2 and 9.3 (Media Cybernetics, USA). The elevation angle ($\theta_{obs}$) was estimated from the ratio of the projection-length of the fin before and after the bending. For mathematical analyses, Matlab versions 2020b and 2021a (Mathworks, USA) were employed.

Development of the mathematical model

Herein, we set a *three-panels-two-torsion springs* object with 100 μm length, 40 μm width, and 1 μm height as a model to simulate the curved bending of the molecular crystal (Figure 1). In this model, we considered the rotational motions of panel-1 and panel-2 and fixed the vertical position of panel-0. The horizontal component of the drag force to the rotational motion of the panels is employed as the thrust for the translational motion. The effect of viscous resistance against translation on the fin-rotation was ignored because the estimated resistance has a small value and low accuracy due to the complexity of hydrodynamics. We simulated three types of motion (Figure 1): in Type-A, a single spring alters the equilibrium point as the result of phase transition; in Type-B, both springs alter the equilibrium point one after the other; in Type-C, both springs alter the equilibrium point simultaneously.

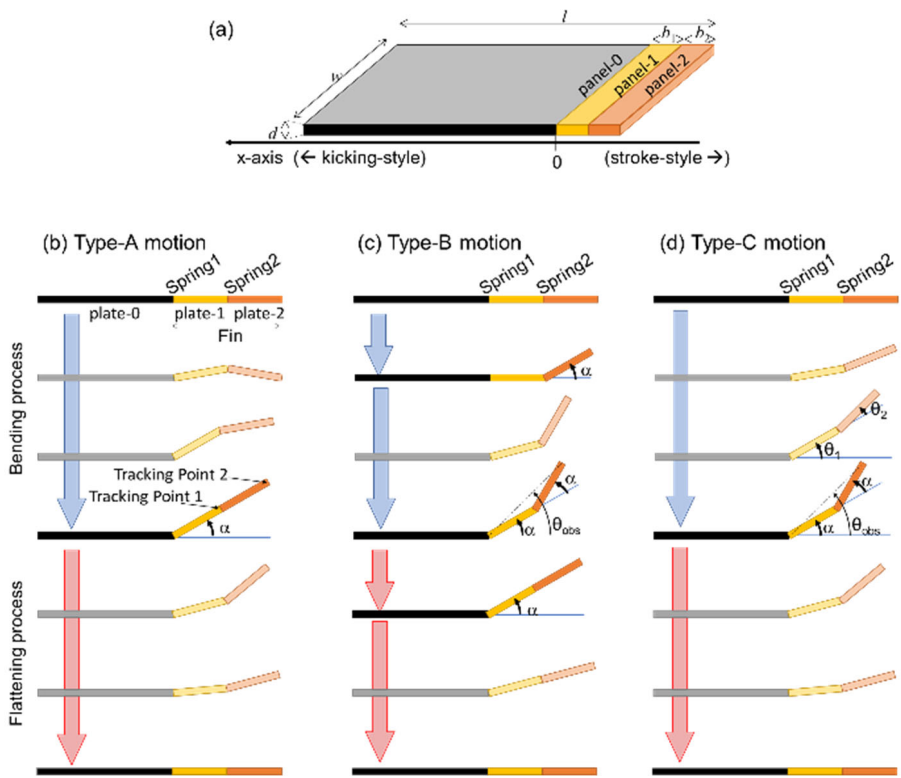

**Figure 1.** (a) Definition of the shape of the object for the three-panel-two-torsion spring model. (b–d) Definition of the angles and points for tracking dynamics and schematic illustration to show the transient orders of the panels' motion for mathematical analysis. (b) Type-A motion, where the balanced angle of spring-1 shifts from 0 to α to trigger the bending process, and from α to 0 to start the flattening process. (c) Type-B motion, in which the bending process occurs via the shift of the balanced angle of spring-2 from 0 to α, followed by the shift of the balanced angle of spring-1 from 0 to α. For the flattening process, the balanced angle of spring-2 shifts from α to 0, and then the balanced angle of spring-1 shifts to 0. (d) Type-C motion, in which the bending process occurs via the simultaneous shift of the balanced angles of both spring-1 and spring-2 from 0 to α, and the flattening process occurs via the shift of both springs from α to 0.

The swimming motion can be described by two equations of motion (EOM). The first EOM is for rotation of the panel (Eq.1):

$$I \frac{d^2}{dt^2} \theta = T_s - T_{Drot} \quad (1)$$

Where $I$ is the inertia moment of the rotating panel, $\theta$ is the elevation angle of the panel as seen from the next panel on the body side, $T_s$ is the torque working on the panel caused by the torsion spring, and $T_{Drot}$ is the torque derived from the drag force working on the panel while flipping. The second EOM is for translation of the set of panels (Eq. 2):

$$m \frac{d^2}{dt^2} x = Th_{transl} - D_{transl} \quad (2)$$

Where $m$ is the mass of the set of panels, $x$ is the position, $Th_{transl}$ is the thrust for the translation obtained as the drag force for the panel's rotation, and $D_{transl}$ is the drag force working on the set of panels against the translational motion.

For simplicity, we assumed that the drag force ($D$) was the linear sum of the drag on the cross-section and that on the shear plane (Eq. 3):

$$D = \frac{1}{2} \rho\, C_d\, S_d\, v_\perp^2 + \frac{1}{2} \rho\, C_f\, S_f\, v_\parallel^2 \quad (3)$$

Where ρ is the mass density of the fluid, $C_f$ and $C_f$, and $S_d$ and $S_f$ are the drag coefficients and areas for the cross-sectional and side-sectional faces of the object, respectively, and $v_\perp$ and $v_\parallel$ are the speeds of the object perpendicular to and parallel to the rotation direction, respectively. According to Azuma[25], $C_d$ and $C_f$ for rectangular objects with low Reynolds number dynamics were reported as (10.9/Re) / (0.96-log Re), and (4 π / Re) / (3.2-2.3 log Re), respectively; to calculate the Reynolds number, Re = $\frac{v L}{\nu}$, ν is the kinematic viscosity of the fluid, $L$ is characteristic length, and $v$ is velocity. Therefore, in our numerical analysis for the object with 40 μm characteristic length, we employ $Z_d$/Re ($Z_d$ = 8) as $C_d$ for rotation, $Z_d$'/Re ($Z_d$' = 2) as $C_d$ for translation, and $Z_f$/Re ($Z_f$ = 1) as $C_f$, because the means of angular velocities and translation velocity were expected to be $10^2$ rad/s (Re = $10^{-1}$~$10^0$) and $10^1$ μm/sec (Re = $10^{-4}$~$10^{-3}$) level, respectively. In addition, we regarded the densities of the fluid (ρ) and the object as $10^3$ kg/m$^3$ and the kinematic viscosity of the fluid (ν) as $10^{-6}$ m$^2$/s. We did not consider additional mass issue in the model.

Next, we built EOM for the rotary motion of each panel.

EOM of panel 2 is presented using maximal elevation angle ($\alpha_2$) as:

$$I = \frac{1}{3} m_2 b_2^2 \tag{4}$$

$$T_s = -k_2(\theta_2 - \alpha_2) \quad \text{for outbound step (bending step)} \tag{5}$$
$$T_s = -k_2(-\theta_2) \quad \text{for inbound step (flattening step)} \tag{6}$$

$$T_{Drot} = \frac{1}{2} \rho v Z_d \left(\frac{1}{2} b_1 \dot{\theta}_1 \cos \theta_2 + \frac{1}{3} b_2 (\dot{\theta}_1 + \dot{\theta}_2)\right) b_2^2 w / L \tag{7}$$

EOM of panel 1 is presented using maximal elevation angle ($\alpha_1$) as:

$$I = \frac{1}{3} m_1 b_1^2 + \frac{1}{3} m_2 (b_1^2 + 3 b_1 b_2 \cos \theta_2 + 3 b_2^2) \tag{8}$$

$$T_s = -k_1(\theta_1 - \alpha_1) \quad \text{for bending step} \tag{9}$$
$$T_s = -k_1(-\theta_1) \quad \text{for flattening step} \tag{10}$$

$$T_{Drot} = \frac{1}{2} \rho v \left(\frac{1}{3} Z_d b_1 + Z_f b_2 \sin^2 \theta_2\right) b_1^2 \dot{\theta}_1 w / L \tag{11}$$

Where $w$ is the width of the panels, $m_i$, $b_i$ are the mass and the length of panel-i, respectively, and $\theta_i$, and $k_i$ are the elevation angle and torsion spring constant of spring-i, respectively.

The parameters for shape and density of the panels were determined from observation. Significant bending and flattening motions were mostly completed within the order of several milliseconds or several ten milliseconds, and the fact that any significant damping vibration of crystals after bending or flattening was not observed indicated that $k_i$ values were around $10^{-15}$ N m. Herein, we regarded that $k_i$ (i=1 or 2) values were proportional to $1/b_i$ and $w$, referring to the relationship between shear stress for rectangular parallelepiped and Young's modulus by Equation 12.

$$k_i = E \frac{w\, d^3}{12\, b_i} \tag{12}$$

Where $E$ is a coefficient of between the stress and strain, and d is the height of the rectangle.

For Type-A motion, we set $\alpha_2$ as 0 and alternated $\alpha_1$. For Type-B motion, $\alpha_2$ and $\alpha_1$ were changed subsequently, and $\theta_i$ was alternated to be constant or variable. For Type-C motion, $\alpha_1$ and $\alpha_2$ were alternated simultaneously, and $\theta_i$ was variable.

For translational motion, we regard that the thrusts are generated by the sum of drag forces against fin rotation. Under this assumption, the acquired forces for horizontal translation by panels 1 and 2 ($Th_1$, and $Th_2$, respectively) are presented as Equations 13 and 14, and the drag force against translation ($D_{transl}$) is presented in Equation 15, in accordance with the equation for the drag force described above (Eq. 3).

$$Th_1 = \frac{1}{4} \rho v Z_d \dot{\theta}_1 b_1^2 w (-\sin\theta_1)/L \tag{13}$$

$$Th_2 = \frac{1}{2} \rho v Z_d \left(b_1 \dot{\theta}_1 \cos\theta_2 + b_2(\dot{\theta}_1 + \dot{\theta}_2)\right) b_2 w (-\sin(\theta_1 + \theta_2))/L +$$

$$\frac{1}{2} Z_f \rho v b_1^2 b_2 w \dot{\theta}_1 \sin\theta_2 \cos(\theta_1 + \theta_2) / L \tag{14}$$

$$D_{transl} = \frac{1}{2} \rho v (Z'_d S_d + Z'_f S_f) \dot{x} \tag{15}$$

Where $Z_f'$ is the coefficient for the relationship between 1/Re and the drag coefficient to the shear face for translational motion. We regarded $S_d$ as the projection area seen from the translational direction, and $S_f$ as the area of projection seen from the vertical direction. Under this assumption, $S_d$ and $S_f$ are presented as

$$S_d = |b_1 \sin\theta_1 + b_2 \sin(\theta_1 + \theta_2)| w \text{ (while } \theta_1 + \theta_2 \geq 0)$$
$$S_d = |b_2 \sin(\theta_1 + \theta_2)| w \text{ (while } \theta_1 + \theta_2 < 0 \text{ and } b_1 \sin\theta_1 < b_2 \sin\theta_2 )$$
$$S_d = |b_1 \sin\theta_1| w \text{ (while } \theta_1 + \theta_2 < 0 \text{ and } b_1 \sin\theta_1 \geq b_2 \sin\theta_2) \tag{16}$$
$$S_f = (b_0 + b_1 \cos\theta_1 + b_2 \cos(\theta_1 + \theta_2)) w \tag{17}.$$

**Results**

Characterization of crystal swimming styles

The crystal showed self-continuous flipping under continuous irradiation by blue light through a sequence of time-irreversible processes, in which light-triggered isomerization of **1** (Steps 1 and 3, in Figure 2) induced a crystalline phase transition (Steps 2 and 4) which repeatedly progressed without external control. The flipping direction was inherent in each crystal, and we assume that the asymmetrical crystal outline and structure, the space group of which is $P_1$,[26] distinguished the bending direction. Due to its limit-cycle mechanism with morphological change (Figure 2), the crystal stored the received energy and then sharply converted it to mechanical power. As a result, the crystal showed an intermittent swimming motion (Movie S1–S3). In other words, the crystal paddled itself forcefully with the intensive transformation at the phase transition steps (Steps 2 and 4). As shown in Figure 3, instantaneous speed was recorded just after the phase transition, which was observed as a sudden change in brightness, after which the crystal stopped the translational motion immediately. On the other hand, the steps for photoisomerization progressed with a slow deformation of the crystal (Steps 1 and 3).   This progression can be understood as a quasi-static process, indicating that the crystal had a low ability to propel itself.

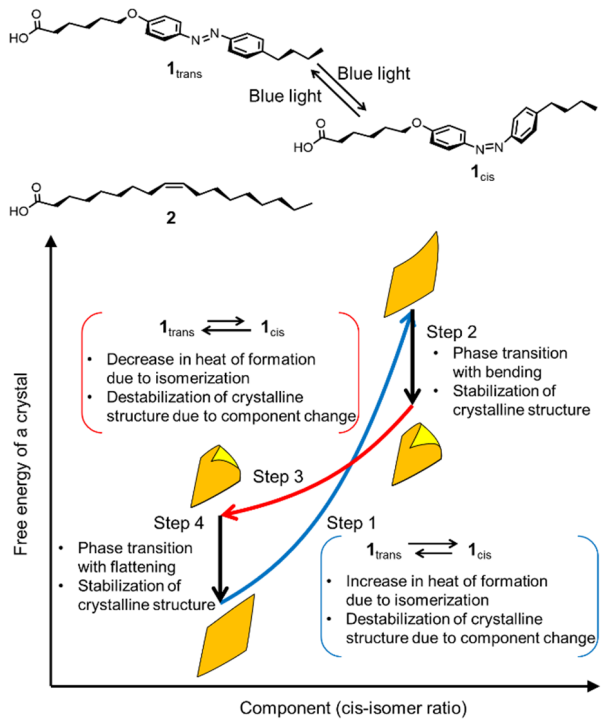

**Figure 2.** Schematic illustration of the energy cycle and explanation of each step for the limit-cycle transition of the self-oscillatory flipping crystal.

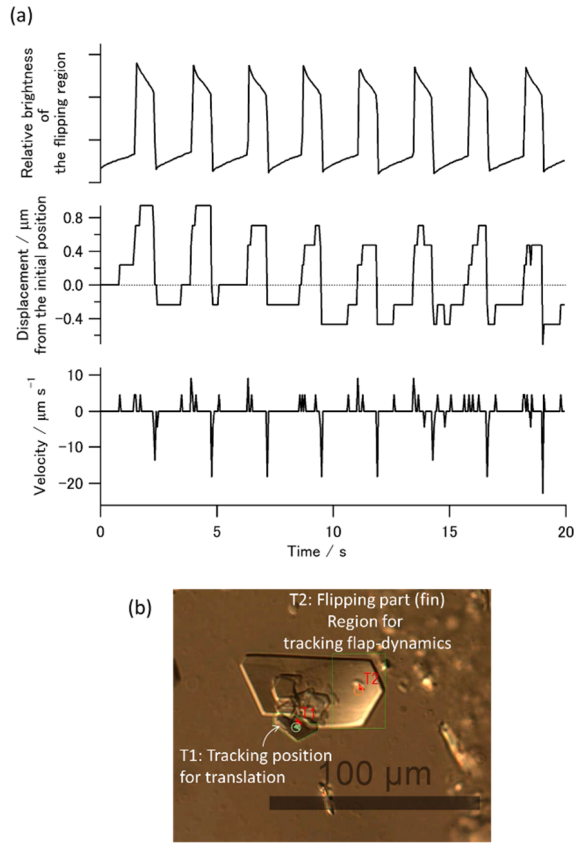

**Figure 3.** (a) Changes in brightness, displacement, and speed of a self-flipping crystal over time. (b) Microscopic image of the crystal with markers indicating the location- and brightness-tracking points used to measure translational motion and fin flipping, respectively. These data were obtained from Movie S3.

Swimming efficiency depended on individual crystals (Figure 4): if the area and elevation angle of the flipped part (the "fin") were larger, then the instantaneous speed was greater. Swimming direction also depended on individual crystals. Many crystals swam in a "stroke" style, with the fin in front (Movie S2); some swam with a "kicking" style in which the fin trailed behind (Movie S1).; others swam in a "sidestroke" style, with the fin off to one side. The crystals that swam in a kicking style tended to lift their large, flat plane fins higher. However, this trend alone did not determine swimming style. Crystals swimming in the stroke style had various-sized fins and elevated them by various angles. In this study, the sidestroke swimmers were classified into either stroke or kicking style, depending on which way their orientation was biased.

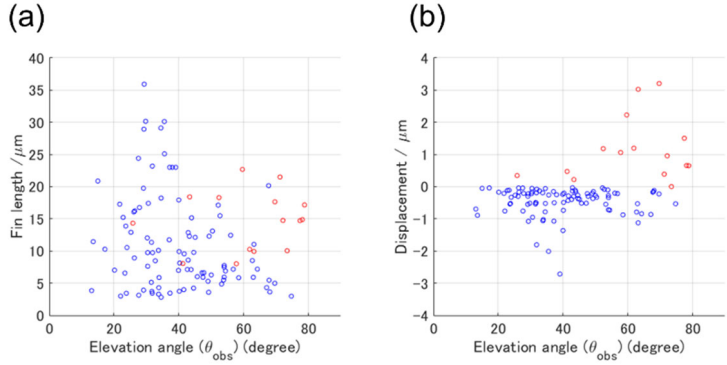

**Figure 4.** Distribution of swimming styles for the observed self-propelling crystals and their correlation with (a) fin length and elevation angle ($\theta_{obs}$) and (b) one-stroke displacement and elevation angle. The color red indicates that the crystal swam with a kicking style, the displacements of which are defined as positive values; the color blue indicates that the crystal swam with a stroke style, the displacements of which are defined as negative values.

Performance of the model

Table 1 shows the calculated results for the motion of an object with a total body length of 100 μm, each fin-length ($b_1$ and $b_2$) of 10 μm, a width of 40 μm, and a height of 1 μm, under conditions of $E = 3 \times 10^3$ Pa and a maximum elevation angle ($\alpha$) of 30°. Using these parameters, the observed elevation angles ($\theta_{obs}$) were 30°, 45°, and 45° for Type-A, -B, and -C motions, respectively. For Type-A and Type-C motions, the object displayed a kicking motion; for Type-B it showed a stroke motion, a result of the backward and forward steps in each process. Figure 5 and Figure 6 illustrate the time course behavior of each model. The differences in fin endpoint trajectory when comparing outbound and inbound routes illustrate that the object's motion is not completely reciprocal. The objects showed unidirectional swimming through an entire cycle of fin-flipping *in-silico* with the assumption that vertical motion of the main body was restricted.

**Table 1.** Calculated displacements of a three-panels-two-torsion springs model via Type-A, Type-B, or Type-C one-round flipping motion.

| Flipping motion | Swimming distance via single flipping / μm | | |
| --- | --- | --- | --- |
| | Outbound | Inbound | Round |
| Type-A | -0.5 | + 3.0 | +2.5 (kicking style) |
| Type-B | -4.6 | + 3.1 | -1.5 (stroke style) |
| Type C | -2.8 | + 4.5 | +1.8 (kicking style) |

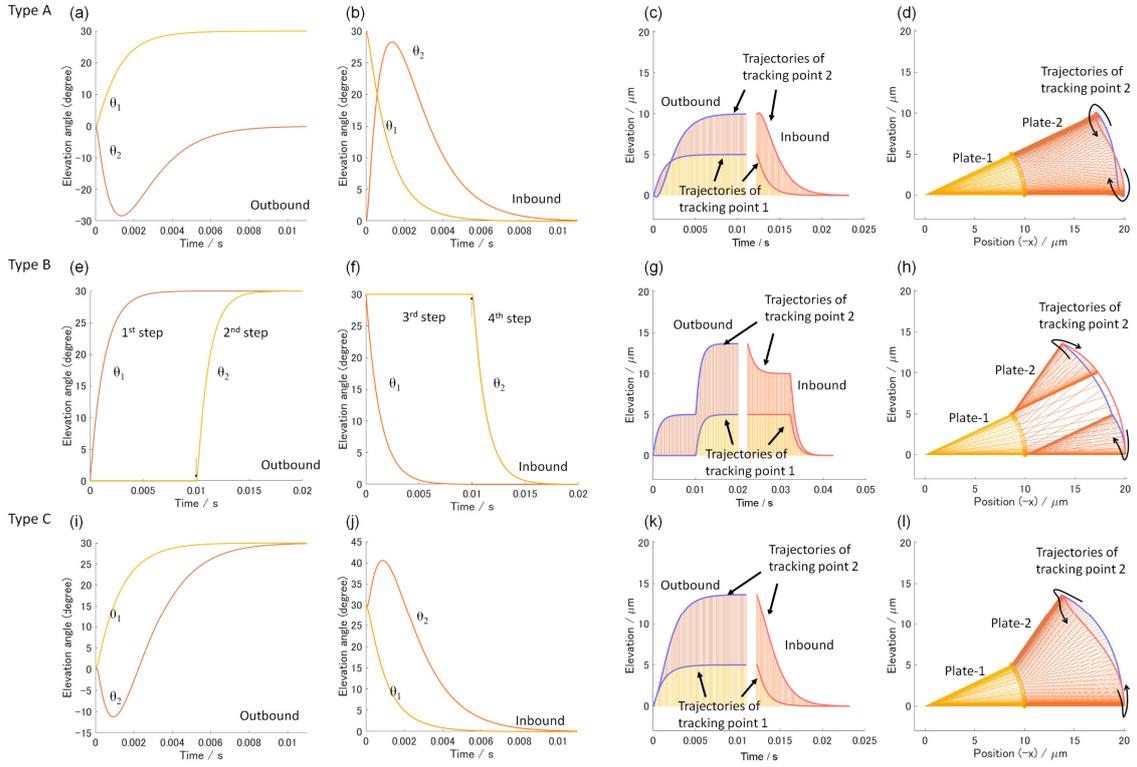

**Figure 5.** Examples of the rotational time-course behavior of panel-1 and panel-2 with 10 μm length each, calculated with Type-A (top), Type-B (middle), Type-C (bottom) motion using $E = 3 \times 10^3$ Pa and $\alpha = 30°$: (a, e, i) shows angular change of each panel during the bending motion (inbound process), (b, f, j) shows angular change during the flattening motion, (c, g, k) shows elevation of the panel endpoints, and (d, h, l) shows a side view of the panels to illustrate the trajectory of the panel-2 endpoint throughout the entire time-course.

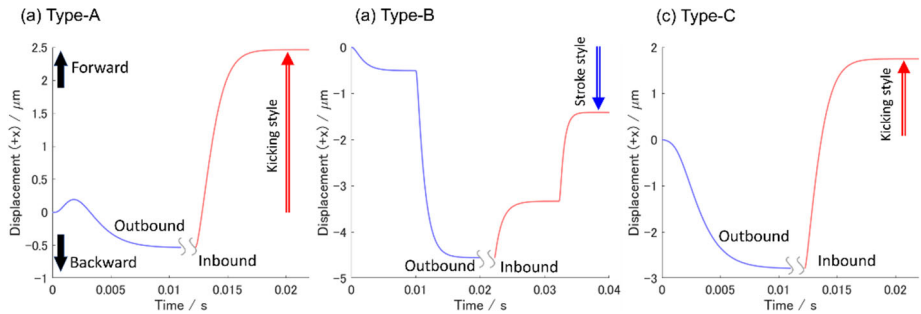

**Figure 6.** The relative translational shift in object positions during one cycle of flipping motion, calculated with (a) Type-A, (b) Type-B, and (c) Type-C motion, using the parameters of $E = 3 \times 10^3$ Pa, $b_1 = b_2 = 10$ μm, and $\alpha = 30°$.

We also solved the equations under various other conditions and calculated the displacements and values of forces and impulses for each. If each maximum elevation angle (α) was less than 60°, the Type-A model resulted in a kicking-type motion, and the Type-B model resulted in stroke motion, independent of the ratio of fin-length. Hereafter we consider only Type-C motion which, like the actual crystals, showed both kicking-type and stroke-type motion,.

Figure 7 shows contour plots for the estimated one-cycle flip displacements of an object with Type-C motion of the same size as the object shown above, calculated with variable fin-length ratio (FR = $b_2/(b_1+b_2)\times 100\%$) and elevation angles. Swimming styles shifted at FR = 41%. For larger FRs, the rotation of panel-2 delayed that of panel-1, and for smaller FRs the rotation of panel-1 delayed that of panel-2 (Figure 8). When FRs were near 41%, the angular velocities of panel-1 and panel-2 were similar to each other. In this case, the panel behaved as a two-panel-one-torsion spring object and showed no net displacement, as predicted by the scallop theorem. The maximum displacement for kicking style was recorded as 3.1 μm at FR = 59% and α = 39° ($\theta_{obs}$ = 62°), and the maximum displacement for stroke-style motion was recorded as 1.2 μm at FR = 27% and α = 40° ($\theta_{obs}$ = 50°). The impulses calculated during bending and flattening for these conditions are shown in Table 2.

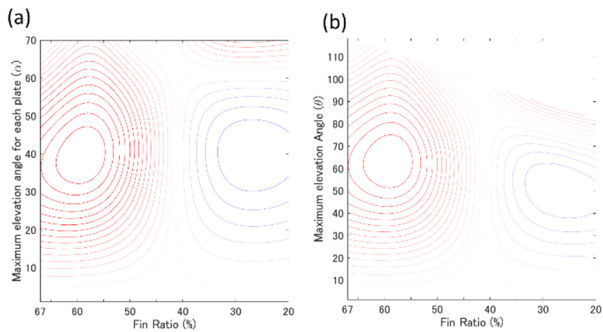

**Figure 7.** Contour plots to present the expected distribution of swimming styles and distances of a 20 μm fin-length flipping object with various elevation angles (α for (a) and θ for (b)) and with various fin ratios (FRs), calculated based on Type-C motion. Red indicates that the object showed kicking motion, while blue indicates that the object showed stroke-style motion.

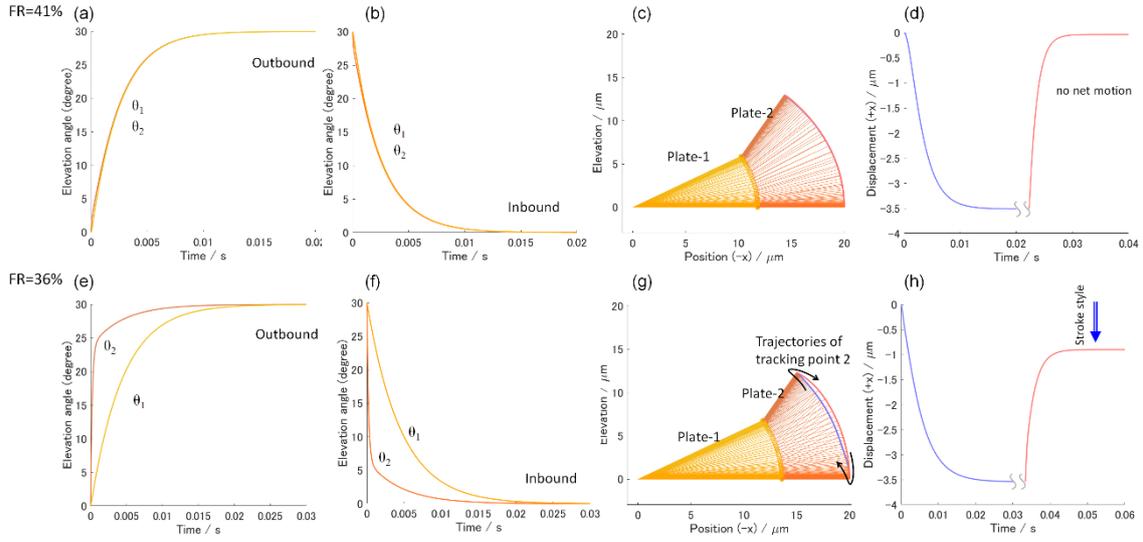

**Figure 8.** Rotational behavior of objects with fin-length ratios of 41% (upper) and 36% (lower). (a, e) and (b, f) are the time-courses of elevation angles of the panels while undergoing bending and flattening, respectively. (c, g) are side-views of panel-1 and panel-2 during a full cycle of the flipping motion; (d, h) are the translation displacements calculated during a full cycle of the flipping motion.

**Table 2.** Calculated impulses applied to the water according to the panels' rotation

| Fin ratio | Outbound (bending) process | | Inbound (flattening) process | |
|---|---|---|---|---|
| | Impulse / N s | Horizontal component of impulse / N s | Impulse / N s | Horizontal component of impulse / N s |
| FR 50% | $5.2 \times 10^{-13}$ | $1.6 \times 10^{-13}$ | $-5.0 \times 10^{-13}$ | $-2.6 \times 10^{-13}$ |
| FR 41% | $4.9 \times 10^{-13}$ | $2.0 \times 10^{-13}$ | $-4.9 \times 10^{-13}$ | $-2.0 \times 10^{-13}$ |
| FR 36% | $4.7 \times 10^{-13}$ | $2.1 \times 10^{-13}$ | $-4.7 \times 10^{-13}$ | $-1.7 \times 10^{-13}$ |

Figure 9 shows the results of model objects with variable fin lengths and elevation angles. The fin length significantly affects the amplitude of displacement and the timescale of motion due to changes in the forces exerted by fin rotation. The value of *E* and thickness affected both the timescale and force strength of the rotation. However, the sum of impulses for a full cycle remained nearly constant as long as the fin did not show damping behavior. Changing the length of the main panel did not affect the rotational behavior of the other panels, but did slightly affect their translational behavior because of a change to the drag on the shear plane. Changing the width affected the strength of forces but did not change the pattern of motion.

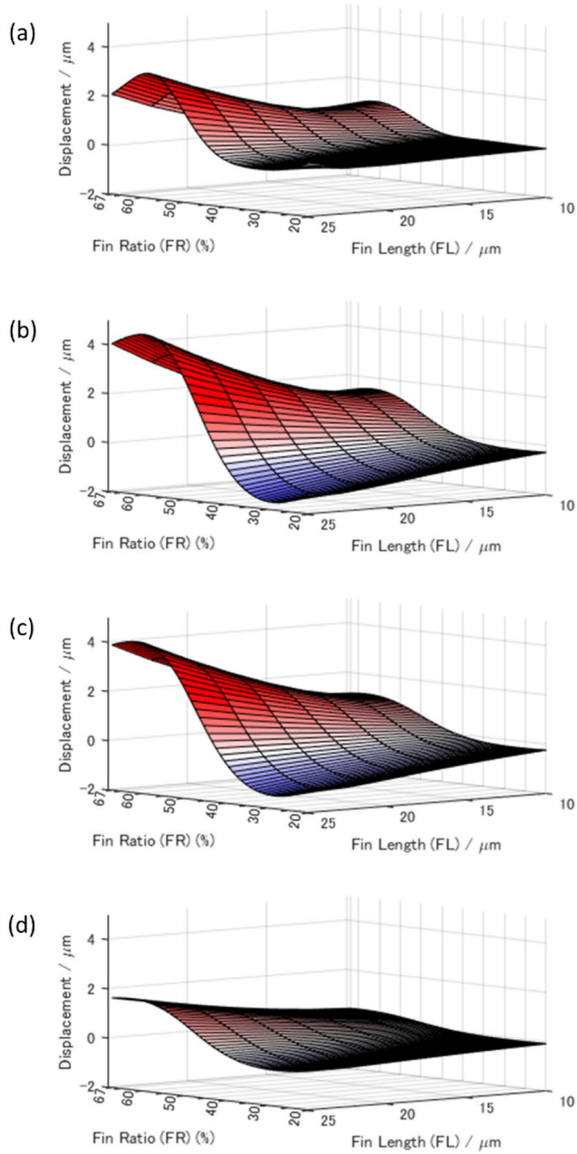

**Figure 9.** Displacement for one-flip motion of a panel calculated with variable fin length (FL = $b_1+b_2$) and fin ratio (FR = $b_2/(b_1+b_2)$) at various maximum elevation angles ($\alpha$). (a) $\alpha = 60°$, (b) $\alpha = 45°$, (c) $\alpha = 30°$, and (d) $\alpha = 15°$.

**Discussion**

The bulk of our comparisons were between different FRs of panel-1 and panel-2. For panels of homogeneous physical properties, the lengths of the panels could be considered equal. Under this assumption, panel-2 would receive a larger drag force to its rotational bending and flattening motion than panel-1. This circumstance would cause the rotation speed of panel-2 to slow, and the motion symmetry to break. As a result, as shown in both Type-C and Type-A motions, the object should show kicking-style swimming. Kicking-style swimming was indeed observed in several crystals. However, most of the crystals showed stroke-style swimming. In terms of the model, stroke style swimming occurred when the rotation of panel-2 preceded that of panel-1. This requirement is satisfied when the phase transition occurs subsequently from spring-2 to spring-1 in Type-B motion, or when the FR was lower than 40% in Type-C motion. In our observations, we were unable to distinguish whether the time delay in the phase transition existed within the tiny crystals. Therefore we could not prove whether or not the crystals swam with Type-B motion. On the other hand, all of the stroke-style crystals had a thick stacked structure at their base side of the fin. Considering that this less-deformable part caused a small FR value, the model for Type-C motion could explain why many crystals swam in this style. The observation that kicking-style crystals had widely-extended planes and thin fins also supports this consideration.

Large displacement with kicking motion can be expected if the FR is larger than 50%, according to the Type-C model shown in Figure 7. However, it is not physically possible to achieve FRs larger than 50% because this would entail the body-side fin being more flexible than the edge-side fin. Figure 10 shows the expected trend between elevation angle and one-flip displacement for crystals with FR $\leq$ 50%. The angle for the peak is expected to be lower in stroke style than in kicking style. The fact that the expected trend is in good agreement with the observed result shown in Figure 4b indicates that type-C motion simulated well the behavior of the crystalline swimmer, although the possibility of a more detailed discussion is limited due to the varying crystal outline. It should be noted that the calculated displacement is somewhat smaller than the observed maximum value, possibly due to an overestimation of $b_0$ and $S_d$.

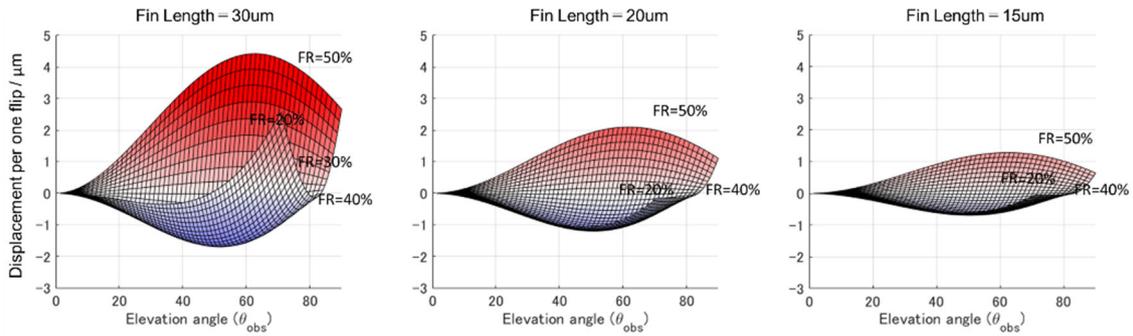

**Figure 10.** Calculated displacements for a one-flip motion of the panel with variable elevation angles ($\theta_{obs}$) and variable FR (20% $\leq$ FR $\leq$ 50%). The fin length of the model object was (a) 30 μm, (b) 20 μm, and (c) 15 μm.

The mechanical properties of the motion were estimated based on the Type-C model, which showed good fit despite our choosing to ignore some of the complexities of hydrodynamics. Impulses to the external fluid for half-rotations (under conditions of α = 30° and FR = 50%) were estimated to be around $5 \times 10^{-13}$ N s due to the accumulation of drag forces and their horizontal components (31% and 52% for bending and flattening, respectively) being transferred to the translational momentum of water flow via inelastic collision. Thus, crystalline rotation leads to translational motion of the water and drag forces acting on the fins, which in turn leads to crystalline translation. And the kinetic energy of the translation immediately and completely transfers to the surroundings. Even if the spring constant is larger, the impulse used for swimming will be the same unless the crystal's shape or deformation style changes. Such mechanics and low efficiency are not unexpected for an object that swims at a low Reynolds number.

In summary, both experimental observation and simplified modeling demonstrated that a self-sustainable bending crystal can swim directionally in a two-dimensional tank. However, the near-reversible flipping motion does not result in swimming with high efficiency. A more efficient motion would be expected from an object that rotates with a large symmetry break, such as in microorganisms that move with a flagellar spiral motion. However, there is a trade-off between the assembled structure of molecules and freedom of movement. Three-dimensional objects have no freedom of motion. On the other hand, two-dimensional objects can perform flip or wave-like motions and swim autonomously, albeit with low efficiency. A one-dimensional assembly may show spiral motion and be an efficient swimmer, but such macroscopic autonomous motion has not yet been synthetically created. Overall, this study shows a basic design for the creation of small, artificial active motion devices.

**Acknowledgements**
Y. K. thanks Dr. Takuji Nakashima (Hiroshima Univ.) for his kind suggestions. Y. K. and K. O. thank Prof. Hideo Kubo (Hokkaido Univ.) for his introductive training for mathematical analysis. The research was supported by JSPS KAKENHI Grant Number JP18H05423 in Scientific Research on Innovative Areas "Molecular Engine" and JP20H04622 in Scientific Research on Innovative Areas "Discrete Geometric Analysis for Materials Design".